\documentclass{article}

\usepackage{PRIMEarxiv}

\usepackage[utf8]{inputenc} 
\usepackage[T1]{fontenc}    
\usepackage{hyperref}       
\usepackage{url}            
\usepackage{booktabs}       
\usepackage{amsfonts}       
\usepackage{nicefrac}       
\usepackage{microtype}      
\usepackage{lipsum}
\usepackage{fancyhdr}       
\usepackage{graphicx}       
\graphicspath{{media/}}     
\usepackage{forest}

\title{The Persian Piano Corpus: A Collection of Instrument-Based Feature Extracted Data Considering Dastgah
}

\author{
  Parsa Rasouli , Azam Bastanfard\thanks{Corresponding author} \\
  Department of Computer Engineering\\
  Islamic Azad University Karaj Branch\\
  Karaj, Iran\\
  \texttt {Parsarasuli@gmail.com} , {Bastanfard@kiau.ac.ir} \\
}

\begin{document}
\maketitle

\begin{abstract} \label{sec: Abstract}
The research in the field of music is rapidly growing, and this trend emphasizes the need for comprehensive data. Though researchers have made an effort to contribute their own datasets, many data collections lack the requisite inclusivity for comprehensive study because they are frequently focused on particular components of music or other specific topics. We have endeavored to address data scarcity by employing an instrument-based approach to provide a complete corpus related to the Persian piano. Our piano corpus includes relevant labels for Persian music mode(Dastgah) and comprehensive metadata, allowing for utilization in various popular research areas. The features extracted from 2022 Persian piano pieces in The Persian Piano Corpus(PPC) have been collected and made available to researchers, aiming for a more thorough understanding of Persian music and the role of the piano in it in subsequent steps.
\end{abstract}

\keywords{Persian Modal Music \and Music data collection \and Iranian piano composers \and Music dataset \and Persian Radif \and Feature extraction}

\section{Introduction} \label{sec: Introduction}
Music, often referred to as a universal language, is an expression of the arts and the diverse cultures of nations. This art form not only plays a significant role as a repository of cultural heritage but also encompasses a wealth of valuable insights and information. The growing interest in the field of music has led to a quest for in-depth analysis, resulting in advancements in research areas such as Music Classification, Acoustic Analysis, and Music Information Retrieval (MIR)\cite{Kassler1966TowardMI}. The domain of Music Information Retrieval is essentially divided into two distinct categories: Content-based and Context-based\cite{INR-042}. In the Content-based\cite{10.1145/860435.860487} field, data and extracted features from audio signals of music pieces are required and can be used in various fields for example explaining a method for finding hidden messages in Persian songs\cite{9050082}. The Context-based\cite{10.1145/2542205.2542206} field, needs comprehensive metadata about the music pieces, which cannot be directly extracted from the music signals. One of the fundamental requirements for advancing this field of research is the availability of extensive information resources with rich metadata for analysis and feature extraction. In the absence of appropriate data, all the strides made in these research domains remain impractical.
On the global stage, particularly in Western music, we observe the creation of vast information resources that greatly assist researchers in this domain. However, the same cannot be said for Eastern music, particularly Persian music. Iran, one of the world's oldest civilizations, has contributed to the spiritual advancement of humanity for well over five centuries and has produced some genuinely unique works of literature and art\cite{caton_1975}. The specific Persian musical mode, known as Dastgah, as well as the unique way of tuning the piano, makes Persian piano music highly valuable. To fully harness the potential of this rich heritage and promote its understanding, it is imperative to establish and gather a substantial amount of data to facilitate research in relevant areas. Unfortunately, there has been very limited activity in gathering Persian music information. This has resulted in constraints in related research fields, highlighting the pressing need to create dedicated information resources. Moreover, there are significant limitations in terms of accessibility and ease of use, given that a large portion of Iranian music data is not available in public repositories. The contributions of the proposed corpus are as follows:

\begin{itemize} \label{sec: Contribution}
\item Addressing Lack of Data in Persian Music Research
\item Promoting Cultural Heritage and Understanding
\item Creation of a Comprehensive Persian Piano Corpus
\end{itemize}

We present an overview of existing efforts in this domain in Section \ref{sec: Related works}, with a focus on Persian modal musical(Dastgah) instruments. However, it is noteworthy that these efforts have primarily omitted a comprehensive representation of Persian piano music.

\section{Preliminaries} \label{sec: Preliminaries}
National music offers a deep awareness of the lives of sages in addition to including moving tales that can bring tears and applause.\cite{chen-2023} 
The availability of music data collections from a nation contributes to the expansion and ease of research in that field, preserves the values and national culture of that country, and significantly aids in achieving more comprehensive research results.

The foundation of Persian art music is a vast collection of tunes called the Radif (row)."The Radif is the principal emblem and the heart of Persian music, a form of art as quintessentially Persian as that nation’s fine carpet and exquisite miniature"(Nettl, 1987)\cite{simms2012mohammad}. During the Constitutional period, Persian musicians Mirza Abdullah(d. 1917) and Ali Naqi Vaziri(b. 1886) dominated the music scene, with Abdullah being the most significant in the area of dastgah music, collecting and classifying melodies from the mid-19th century. Mirza Abdullah's radif passed down through his father, Ali Akbar Farhani, is considered the foundation of the classical Persian music tradition.\cite{caton_1975}
At present, there are seven Dastgah(a traditional Persian musical modal system which is a melody type) and five Avaz (or Naghme which are considered auxiliary dastgah-ha)in the canonical repertoire.\cite{babak_nikzat_2022_7316660}
Vaziri, an Iranian military officer, studied music in Tehran, Paris, and Berlin, aiming to establish a European-style conservatory in Iran. He founded the Madresse-ye Ali-ye Musiqi Conservatory in 1923.\cite{caton_1975}  He Also made great contributions to Persian piano and also a special tuning method for piano which includes Quartertones. Several pieces from students of the Madresse-ye Ali-ye Musiqi Conservatory, such as Javad Maroufi, are also present in this corpus.

The piano's significance in Persian music necessitates in-depth research on this instrument and piano music. However, the availability of relevant data is lacking. Such research can enhance the development and enrichment of Persian music, opening new horizons for artists and researchers. Data collections related to piano music can serve as crucial tools in Persian music research, enhancing knowledge and aiding in scientific and artistic research. This enables a more profound analysis of the piano and piano music.

\section{Review of related works} \label{sec: Related works}
We understand the significance of national music and related research when we delve into the influential history of music. For instance, In \cite{chen-2023} the importance of China’s traditional national music and the need for integrating college music education with national music culture is stated.
In \cite{african} the cultural significance of African Christian music is highlighted, which represents interrelated conceptual and physical movements, central to Christian mission and expansion in Africa.
In \cite{pingle-2023} the human body's response to the Indian classical raga structure in the context of music therapy is investigated. The study highlights the importance of Artificial Intelligence(AI) in providing valuable patient information and recommendations, enhancing researchers' understanding of music therapy's effects on various diseases. A similar method can also be employed to examine the influence of Persian music pieces.
Now, for further research similar to what was mentioned in the first step, there is a need for suitable and comprehensive data collection. Unfortunately, due to the lack of any relevant data related to the Persian piano, in the following section, we mention approximately similar cases in the field to create corpora to facilitate indexing, retrieved music research domain. Music data collections can be categorized based on various criteria. we categorized and explained data collections based on the main purpose of the creation including Size-based, Mode-based, and Task-based approaches.

\subsection{Size-based approach} \label{sec: Size based}
 There are some large-scale corpora where the main point is the huge amount of data that can be used for music indexing and information retrieval, for example, in American pop music, The USPOP2002\cite{10.1162/014892604323112257} is a massive music corpus, containing 706 albums and 8764 tracks from 400 American pop artists. It includes artist and album names, music styles, and genre identification. The corpus is available on three DVDs for researchers, but MFCC-derived features are used to prevent copyright violations.
In another attempt to gather Iranian music, a Large Corpus of Iranian Music\cite{5407527} was published which presents a comprehensive corpus of Iranian music, containing 27496 tracks from 1355 artists sourced from the internet. The corpus also includes artist and album names, as well as the lyrics of 54 percent of songs. The first 20 MFCC files are extracted and saved in HTK format, and the extracted features are then distributed. 
One of the main drawbacks of this data collection method is the lack of attention to the quality of the used pieces and the presentation of appropriate metadata. This is due to the fact that these details have occasionally been ignored in an effort to expand the volume of data, and the lack of appropriate labels for the data limits the datasets' potential applications.

\subsection{Mode-based approach}\label{sec: mode based}
In the mode-based classification method, such as in the case of the Persian Dastgah, limited activities have been carried out, resulting in the creation of some data collections, such as \cite{gujsa98108} in 2013 which classified the Radif using parameters like pitch frequency, kurtosis, and spectral centroid. It gathered data from 1250 musical compositions for Iranian instruments like Zehi, Zakhme, Tar, and Setar from five CDs. The data, aided by Setar and Tar masters, has not been publicly available.

The Maryam Iranian Classical Music Dataset (MICM)\cite{https://doi.org/10.13140/rg.2.2.18688.89602} was created in 2018 with the purpose of analyzing and
categorizing Iranian classical music. MICM contains 3311 music samples, 631 of which include a straw instrument
in the foreground and other instruments in the backdrop. The remaining samples are divided into seven groups,
representing the seven Iranian Classical Dastgah. There is no categorization or metadata for Avaz, and the audio files only include waveform files and labels. In another research\cite{malekzadeh-2019}, This dataset was utilized and AlimNet, an ACGAN, was used to generate categorically generative music.

The Nava \cite{Nava} dataset was published in 2019 and contains 1,786 instrumental performances by 40 Iranian artists, each in one of seven traditional music modes and using one of the five Iranian instruments: Tar, Santoor, Setar, Kamancheh, or Ney. Feature extraction from the music signal involved computing Mel-frequency cepstral coefficients and extracting feature vectors, which were then transformed into a single vector using the feature concatenation method. The Nava dataset was used in research like \cite{ebrat2022iranian} where several deep neural networks were implemented to recognize Iranian modal music in seven Dastgah.

Another instance is the KDC\cite{babak_nikzat_2022_7316660}, a collection of Iranian dastgahi music recordings and data, analyzed using computational methods that was created in 2022. It includes 213 solo recordings by professional musicians, complete performances of 7 dastgah and 5 avaz, and contributions from four musicians. In contrast to MICM, The corpus includes metadata and annotations, including dastgah, guše, artist, instrument, recording type, and duration. This corpus includes the Ney and Setar instruments, and Similar to MICM, it provides music tracks in FLAC format.

Research has been conducted on the Santoor, which is another traditional instrument in Iranian music. In one study\cite{Santoor1}, one of the authors recorded 91 pieces (approximately one and a half hours of music) to obtain the required data and then used computational
analysis to determine the dastgah. In another work\cite{Heydarian2005ADF}, a small dataset consisting of single notes and Santoor melodies has been provided to address the challenge of data scarcity in the field of Iranian music. This dataset aims to assist in mitigating the data shortage for research in Iranian music.

The main problem with the Mode-based approach is the lack of comprehensiveness and completeness in the provided pieces, which results in a limited amount of valuable information related to the presented musical instruments. The reason is due to the fact that all genres of modern Persian music incorporate traditional instruments. Focusing on only a portion of these pieces leads to an incomplete research perspective regarding the instruments.

\subsection{Task-based approach} \label{sec: Task based}
In this approach, scientists are primarily focused on streamlining their research efforts in their respective domains by gathering data relevant to their expertise. Most of the datasets presented in this section are subsets of larger data collections, where specific criteria are applied to select and process the data, such as The Meter2800 dataset\cite{ABIMBOLA2023109736} is a collection of 2800 annotated audio tracks designed for time signature detection. It includes audio files from three popular MIR datasets, GTZAN\cite{tzanetakis-2002}, MagnaTagATune\cite{Magna}, and FMA\cite{defferrard-2017}, as well as additional tracks. The dataset also includes pre-computed features, tempo, and time signature, including Chroma Short-Time Fast Fourier Transform, Root Mean Square, Mel-Frequency Cepstral Coefficients, Spectral Centroid, Spectral Bandwidth, Zero Crossing Rate, and Spectral Rolloff.
This approach is also largely derived from larger data collections and is used specifically for certain research purposes.

\begin{figure}
  \centering
    \begin{minipage}{0.7\linewidth}
    \centering
    \includegraphics[width=\linewidth]{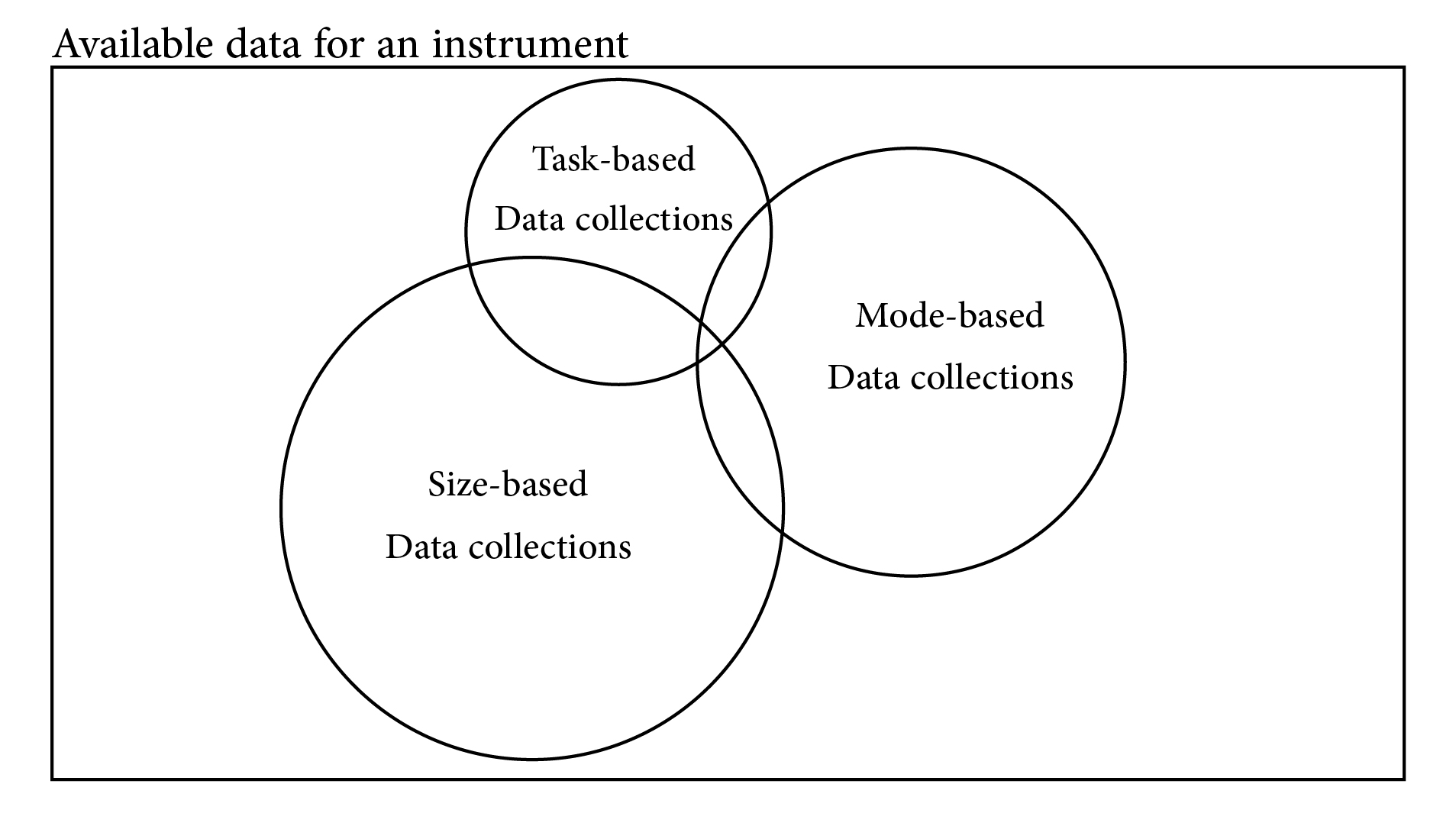}
    \caption{The illustration of existing data collections compared with the whole available data for an instrument.}
    \label{fig:Repre}
    \end{minipage}
\end{figure}

The main challenge and limitation in collecting data under the mentioned categorizations are the domain restrictions and their lack of comprehensiveness. For example, in the \ref{sec: mode based}, we mentioned some data collections related to instruments like the Santoor or the Violin, but the question arises: Are these pieces comprehensive and complete? Are there pieces from these instruments in other musical modes or genres missing? Are these resources sufficient for understanding music produced with these instruments or even Iranian music as a whole? Is there data available for all the instruments used in the music of different nations using this method? Does this method result in missing valuable information about instruments that are less commonly used in these modes?

Researchers often face these challenges when working with limited data collections.
The existing data collections can not cover all the available data for an instrument as shown in \ref{fig:Repre}. To achieve a more comprehensive understanding, it's essential to seek diverse and representative data collections that cover a wide range of musical styles, genres, and contexts, which can better represent the richness and diversity of the music produced with specific instruments and within specific musical traditions.

As the primary instrument under examination, the piano plays a significant role as a mother instrument in music. There is a substantial amount of data available for it in the realm of classical music, such as the Piano Concerto dataset\cite{ozer-2023}. However, a similar situation cannot be claimed for Persian music. As stated in \cite{Vafaeian}, the lack of coherent datasets and Persian music data is the main reason for the ineffectiveness and desirability of results in the field, often hindering their implementation and the development of practical software.

\section{Proposed Method} \label{sec: method}
All the categories mentioned in section \ref{sec: Related works} are areas that have contributed to research and development in the field of music up to this day. However, to address the posed questions and overcome existing limitations, the instrument-based approach must also be utilized. This approach ensures that all aspects of an instrument are thoroughly examined, and pieces in various genres and modes are comprehensively represented. This comprehensive approach allows for not only instrument recognition but also utilization in other areas such as music information retrieval, genre classification, and more. Data collections focusing on a specific instrument, especially in the music of a particular nation, can provide more insights into how changes have occurred over time or reflect cultural variations.

\subsection{Design Objectives} \label{sec: Objective}
Most of the available resources in the field of instrument recognition are focused on Persian traditional instruments such as Tar, Setar, and Santoor, and there is no data related to the piano. Additionally, when it comes to international genres like pop, there is a lack of data regarding Iranian composers, which creates limitations in researching and studying Persian music.

The complete metadata provided in the proposed corpus is another advantage since it includes comprehensive information about the pianist, composer, album, and year. This information is crucial for indexing and information retrieval systems.

Preserving and paying attention to authentic Iranian art has been one of our primary responsibilities. We have strived to make the rich culture of Persian music accessible to everyone for more in-depth exploration and analysis by drawing from Persian piano masters from decades past to the present day.

\subsection{Data gathering} \label{sec: Data gathering}
We examine a few of the music corpora used in research on music indexing and retrieval in section\ref{sec: Related works}. Taking into account the significance of the Iranian piano, as discussed in section \ref{sec: Preliminaries}, we decided to create a Persian Piano corpus and We began to gather audio files from the Internet. We collected a significant number of MP3 and FLAC files of Iranian piano music during our crawls. 
Subsequently, we started deleting duplicate files and files with additional issues, like tracks that are less than 10 seconds in length or format issues. We have some 2022 music tracks at the end. We will explain multiple aspects of PPC corpus in the remaining sections of this corpus. We tried to gather pieces from the different time periods, taking into account the rich history of piano music in Iran. The absence of a primary source for preserving and maintaining such works led us to encounter various resources with varying quality. However, this diversity contributed to the breadth of the provided data. One of the main challenges in collecting this corpus was related to copyright laws and the unavailability of audio files for the pieces. After assessing similar resources and considering the essential requirements of researchers in this field, we decided to provide features extracted from the pieces. Since re-creating the pieces with suitable quality based on these features is not practical, PPC corpus does not violate copyright laws. The varying sizes of the pieces available on the internet were another major issue. However, considering the various aspects of this corpus, such as its focus on comprehensiveness and the preservation of Persian music heritage, we tried to minimize the removal of pieces. Instead, we provided relevant labels like solo performances or Additional information alongside the corpus.

\subsection{Feature Extraction}\label{sec: Features}

It is necessary to perform some sort of feature extraction as music files' waveforms cannot be used directly for indexing and retrieval.\cite{5407527} Also the fact that music track extracts can be shared without breaking copyright is another significant point to remember.\cite{10.1162/014892604323112257}
As mentioned in \ref{sec: Related works} section, some of the prior works like USPOP2002 and LCIM utilized features extracted from MFCCs in their corpora, and in some other cases, such as MIR-2800, we can see the presence of other features, such as Root Mean Square, Spectral Centroid, Spectral Bandwidth, Zero
Crossing Rate, etc. The features we extract are chosen for their usefulness in understanding problems and proposing optimal solutions for music-recommended applications.\cite{jitendra-2020} We also adopted a similar approach to prepare most of the necessary features for use in various research domains.
Transferring just derived features can help to circumvent the major barriers to sharing or distributing huge music collections, which should also result in a reduction of bandwidth requirements.\cite{10.1162/014892604323112257} Providing extracted features combined with the metadata enables researchers to use new methods like combining audio information with meta-information, such as artist or genre. It uses spectrum and periodicity histograms to describe timbre and rhythm.\cite{10.1162/014892604323112248} A third type of organization, not derived from audio analysis, is also introduced. The music is organized on a map, with similar pieces located close to each other, using an Islands of Music metaphor.
Researchers utilized low-level musical features and factors to model musical domains, describing music moods and music-related situations based on time, location, and subject.\cite{rho-2011}
In another attempt various music genres are examined to determine the suitable content for radio broadcasting. One of the main points mentioned is the necessity of extracting different features as an initial step which is one of the main steps in the creation of this music data corpus.\cite{7288794}
As a result, having resources of features is one of the starting points for a wide range of research.

In this section, we will provide details on these aspects.

The process whole begins by loading each audio track from the local storage of tracks into the Python environment to analyze and extract information from the audio signal.

Afterward, spectrograms are extracted to provide insights into the frequency components and their evolution over time. Then The following audio features are extracted using Librosa\cite{mcfee_2023_8252662}:

\begin{itemize}
\item Chroma Short-Time Fast Fourier Transform (STFT): Chroma features represent the energy distribution of pitch classes, providing information about the tonal content of the music.

\item Root Mean Square (RMS): RMS measures the energy in an audio signal, offering insights into loudness.

\item Mel-Frequency Cepstral Coefficients (MFCCs): MFCCs are coefficients that represent the short-term power spectrum of a sound signal, resembling the human auditory system's response.

\item Spectral Centroid: Spectral centroid indicates where the center of mass of the spectrum is located, conveying information about the brightness of the sound.

\item Spectral Bandwidth: Spectral bandwidth defines the width of the spectral envelope and is related to the timbral texture of the music.

\item Zero Crossing Rate: Zero crossing rate calculates the rate at which the audio signal changes sign, revealing information about its noisiness.

\item Spectral Rolloff: Spectral rolloff signifies the frequency below which a particular percentage of the total spectral energy is contained.
\end{itemize}

The extracted features of music include chroma features, which capture tonal content, Root Mean Square (RMS), which represents loudness, and Mel-Frequency Cepstral Coefficients (MFCCs), which are used for speech and audio processing tasks. Spectral centroid is valuable for understanding the tonal characteristics of audio signals, while spectral bandwidth provides insights into the timbral texture of music. Zero crossing rate indicates audio noisiness, which can be applied in tasks like speech analysis and audio quality assessment. Spectral roll-off helps identify the spectral characteristics of audio signals, contributing to tasks like genre classification and audio effect detection. These features are versatile and can be used for various music-related tasks, such as music genre classification, instrument recognition, acoustic analysis, and music information retrieval. The combination of these features enables accurate genre classification based on tonal and spectral characteristics. Additionally, these features can be used for acoustic analysis, allowing users to search for audio tracks based on their acoustic characteristics.

\section{Result} \label{sec: Result}
The Persian Piano corpus\cite{DVN/YY7SVD_2023}(accessible from: \hyperlink{https://doi.org/10.7910/DVN/RQSQY8}{https://doi.org/10.7910/DVN/RQSQY8}), featuring valuable pieces from the earliest Persian piano music composers to contemporary prominent figures in this field, has been meticulously collected. Through necessary analyses, it has been compiled in a unified format and made publicly accessible. This corpus aims to empower researchers to delve into specialized investigations, contributing to new insights and discoveries within the realm of Persian piano music. In the following sections, we will present some features and statistics related to this corpus.

\begin{table}
  \centering
  \begin{tabular}{|l|r|r|r|r|r|}
    \hline
    Composers & Tracks & Total Length & Albums & Piano Solos & Labeled\\
    \hline
    Fariborz Lachini & 1047 & 74:58:06 & 1047 & 1011 & -\\
    Morteza Mahjoubi & 159 & 5:34:49 & 159 & 148 & 159\\
    Javad Maroufi & 152 & 18:51:14 & 152 & 115 & 26\\
    Moshir Homayoun Shahrdar & 113 & 2:42:05 & - & 95 & 98\\
    Hossein Ostovar & 93 & 6:27:55 & - & 48 & 89\\
    Fakhri Malekpour & 71 & 2:31:10 & - & 71 & 71\\
    Arash Behzadi & 48 & 4:34:52 & 48 & 41 & -\\
    Andre Arezoomanian & 47 & 3:09:59 & 47 & 20 & -\\
    Saman Ehteshami & 43 & 3:17:43 & 43 & 20 & 7\\
    Mohsen Karbassi & 37 & 2:49:00 & 37 & 37 & -\\
    Anoushiravan Rouhani & 33 & 2:55:29 & 33 & 24 & 2\\
    Alireza Lachini & 22 & 1:45:40 & 22 & 22 & -\\
    Saeed Deyhemi & 18 & 0:46:42 & 18 & 18 & -\\
    Rooh Angiz Rahgani & 17 & 0:56:13 & 17 & 17 & -\\
    Shahrdad Rouhani & 16 & 1:03:38 & 16 & 16 & -\\
    Ardeshir Rouhani & 15 & 1:36:09 & 15 & 1 & 6\\
    Pooyan Azadeh & 14 & 1:04:34 & 14 & 14 & 6\\
    Pouya Nikpour & 14 & 1:07:51 & 14 & 14 & -\\
    Mohammad Mahdi Abolhoseini & 14 & 0:50:53 & 14 & 13 & -\\
    Ahmad Abedi & 10 & 0:45:41 & 10 & 10 & -\\
    Shahriar Rohani & 9 & 0:43:02 & 9 & 0 & -\\
    Maziar heidari & 8 & 0:22:30 & 8 & 8 & -\\
    Arman Nahrvar & 8 & 0:36:37 & 8 & 8 & -\\
    Pezhman Mosleh & 8 & 0:46:54 & 8 & 8 & -\\
    Mohammad Sangian & 6 & 0:16:20 & 6 & 6 & -\\
    Grand Total & 2022 & 140:35:06 & 1745 & 1785 & 464\\
    \hline
  \end{tabular}
  \caption{Statistics of Persian Piano Music Composers in our corpus}
  \label{tab: composers}
\end{table}

\subsection{Composers and Tracks}
\begin{figure}
  \centering
  \begin{minipage}{0.48\linewidth}
    \centering
    \resizebox{\linewidth}{!}{\includegraphics{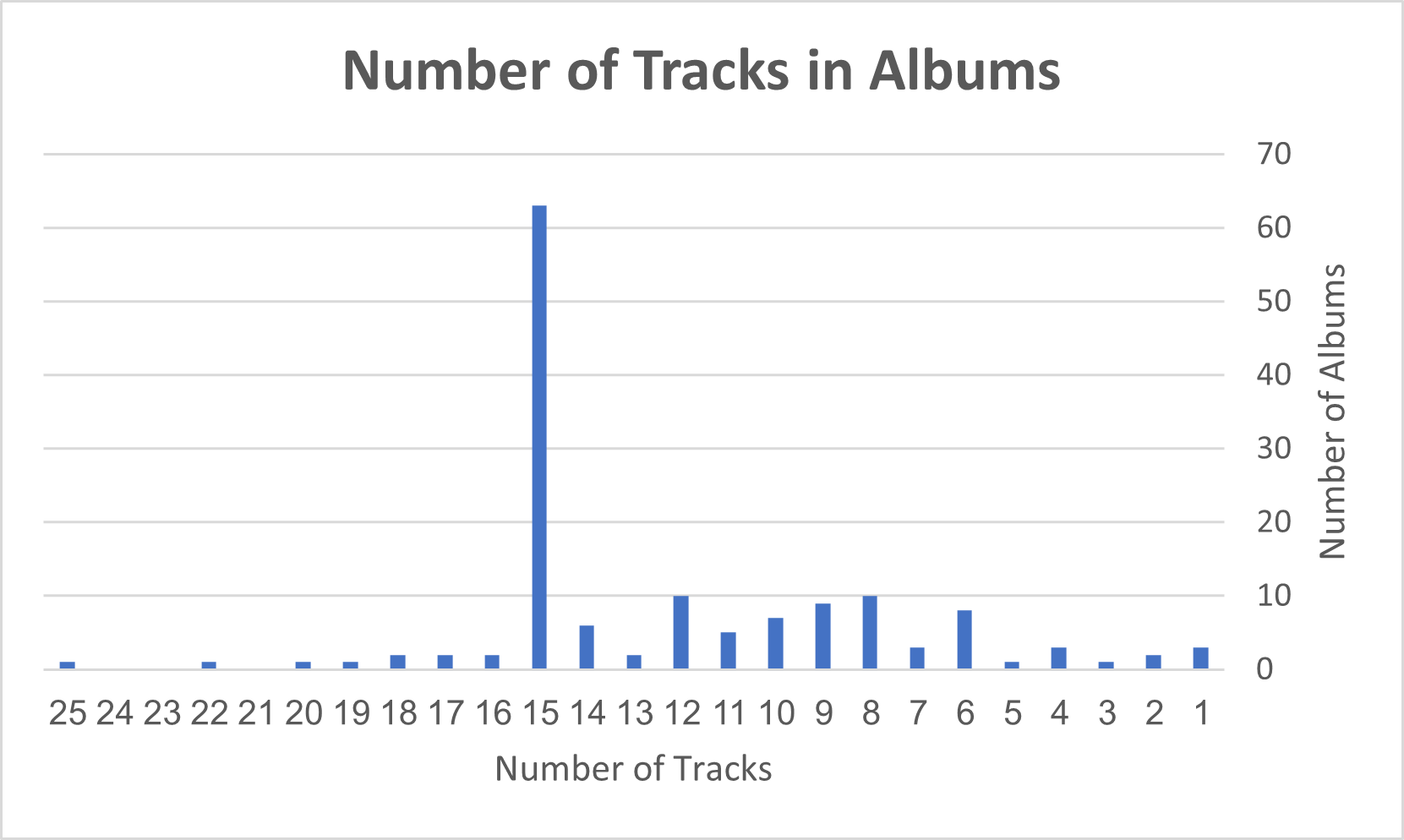}}
    \caption{The distribution of the number of the Piano tracks in albums.}
    \label{fig:NTA}
  \end{minipage}
  \begin{minipage}{0.48\linewidth}
    \centering
    \resizebox{\linewidth}{!}{\includegraphics{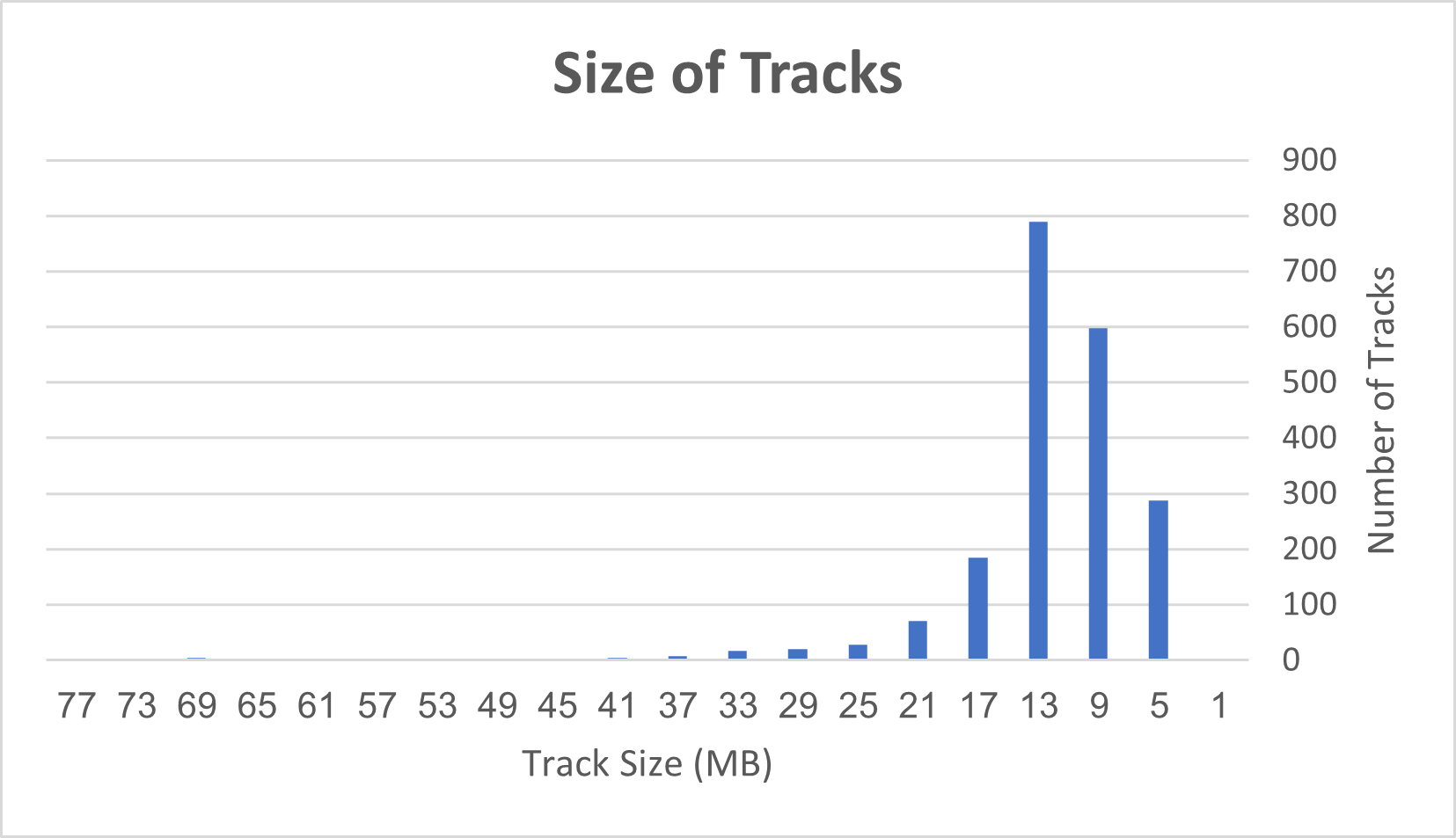}}
    \caption{The distribution of tracks file sizes.}
    \label{fig:SoT}
  \end{minipage}
\end{figure}

Among the 2022 pieces collected, there are 25 music composers along with 26 piano performers. This means that some of the pieces, especially older compositions, have been performed by various pianists. Within the list of mentioned composers in this corpus, the names of Iranian piano pioneers such as Morteza Mahjoubi stand out.
The scope of the pieces available from each composer is visible in Table \ref{tab: composers}, where the largest number of pieces is attributed to Fariborz Lachini, who has had a significant influence on Iranian piano music. 
Among the 2022 pieces collected in this corpus, 1745 of them are from music albums. The number of pieces within the albums in this corpus varies between 1 and 25, and you can view the distribution of the number of pieces in albums in figure \ref{fig:NTA}.

\subsection{Length and Size}
As previously mentioned, pieces shorter than 10 seconds have been removed from this corpus, and the maximum length of the pieces is 20 minutes. The distribution of piece duration is illustrated in figure \ref{fig:LT}.
In this corpus, file sizes vary due to the use of different file formats, quality levels, and the duration of the pieces. The smallest file sizes are 10, and the largest file sizes are 20. You can observe the distribution of different file sizes in figure \ref{fig:SoT}.

\subsubsection{Format and Bitrate}
The use of various formats in the PPC corpus, including MP3 and FLAC, has created a valuable range of pieces with different qualities and characteristics. Additionally, the presence of original and older pieces contributes to the observation of different bitrates. The diversity of formats, qualities, and bitrates adds richness to the corpus, making it suitable for various research and analysis purposes.

\subsection{Labels}
\begin{figure}
  \centering
  \begin{minipage}{0.4\linewidth}
    \centering
    \begin{tabular}{|l|r|}
      \hline
      Category & Count \\
      \hline
      Avaz Abuata & 36 \\
      Avaz Afshari & 51 \\
      Avaz Bayat-e Esfahan & 53 \\
      Avaz Bayat-e Tork & 45 \\
      Avaz Dashti & 40 \\
      Dastgah Chahargah & 34 \\
      Dastgah Homayun & 44 \\
      Dastgah Mahur & 33 \\
      Dastgah Nava & 25 \\
      Dastgah Rast-Panjgah & 22 \\
      Dastgah Segah & 45 \\
      Dastgah Shur & 36 \\
      Grand Total & 464 \\
      \hline
    \end{tabular}
    \caption{Category Counts}
    \label{tab:category-counts}
  \end{minipage}%
  \begin{minipage}{0.6\linewidth}
    \centering
    \fbox{\includegraphics[width=\linewidth]{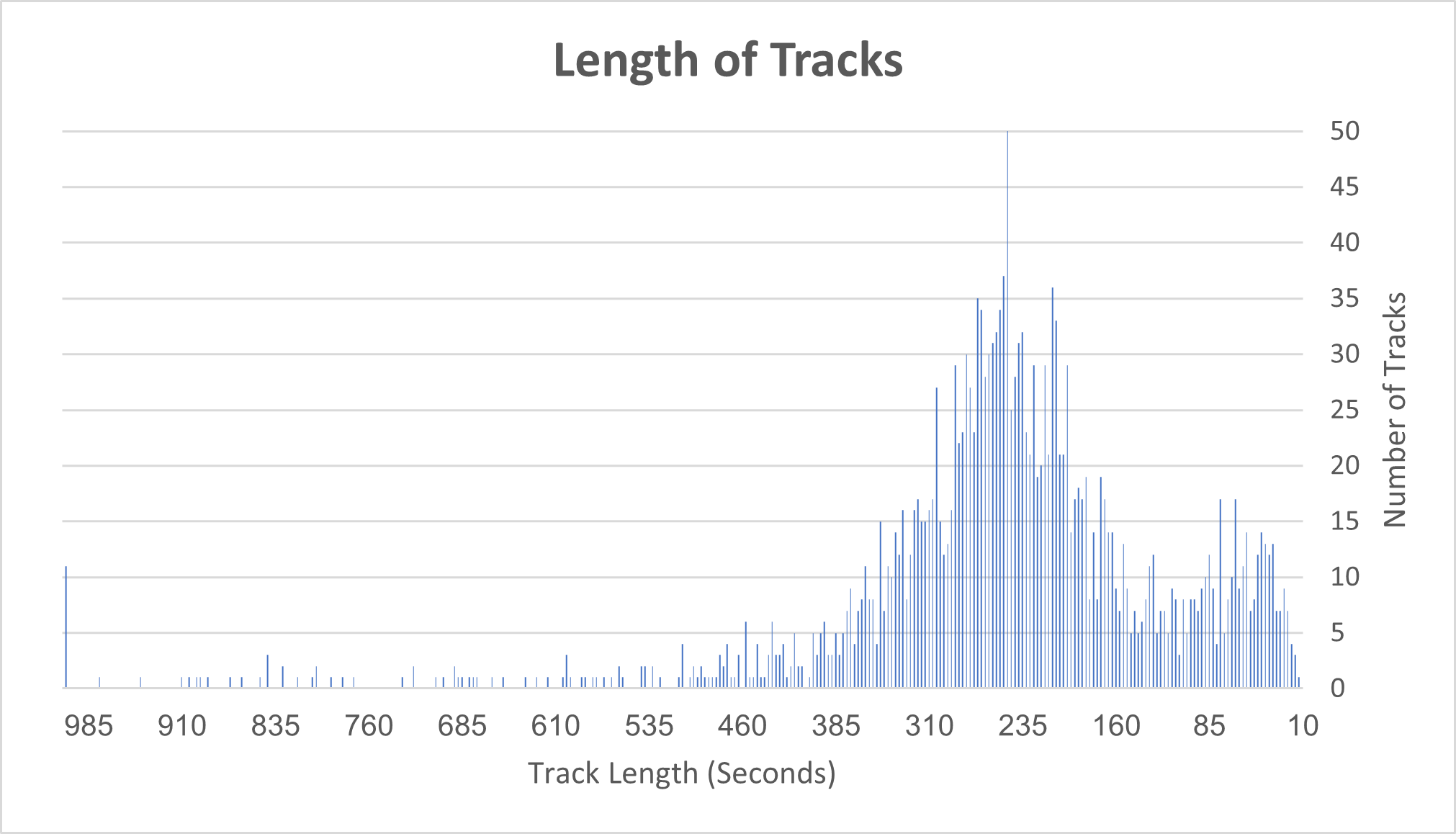}}
    \caption{The distribution of the length of the Piano tracks.}
    \label{fig:LT}
  \end{minipage}
\end{figure}

For each of the pieces in this corpus, there are three available information fields:
\begin{itemize}
\item Radif Label: The labels of the pieces are based on 7 Dastgah and 5 Avaz. These labels indicate the musical mode of the respective piece and are suitable for use in a system for instrument detection and other research purposes. The number of pieces for each label is specified in table \ref{tab:category-counts}.

\item Piano Solo: This field serves as an indicator for the instruments used in each piece. A value of 1 signifies a solo piano, while a value of 0 indicates the presence of multiple instruments or vocal modes in the piece. Among the 2022 pieces in this corpus, 1785 of them are solo piano pieces.
Vocals are included in some unique compositions as well; these can be isolated using related corpora like \cite{10.1007/978-3-642-11301-7_30} as well as techniques like \cite{9349583}, which is peculiar to the Persian singing voice.

\item Additional: Supplementary information about the pieces is provided, including details such as the names of other performers (in cases where the piece is not solo), information about the originality and age of the piece, and mentions of Iranian Tuned piano.

\end{itemize}

\begin{figure}
  \centering
  \begin{minipage}{0.5\linewidth}
    \centering
    \resizebox{\linewidth}{!}{\includegraphics{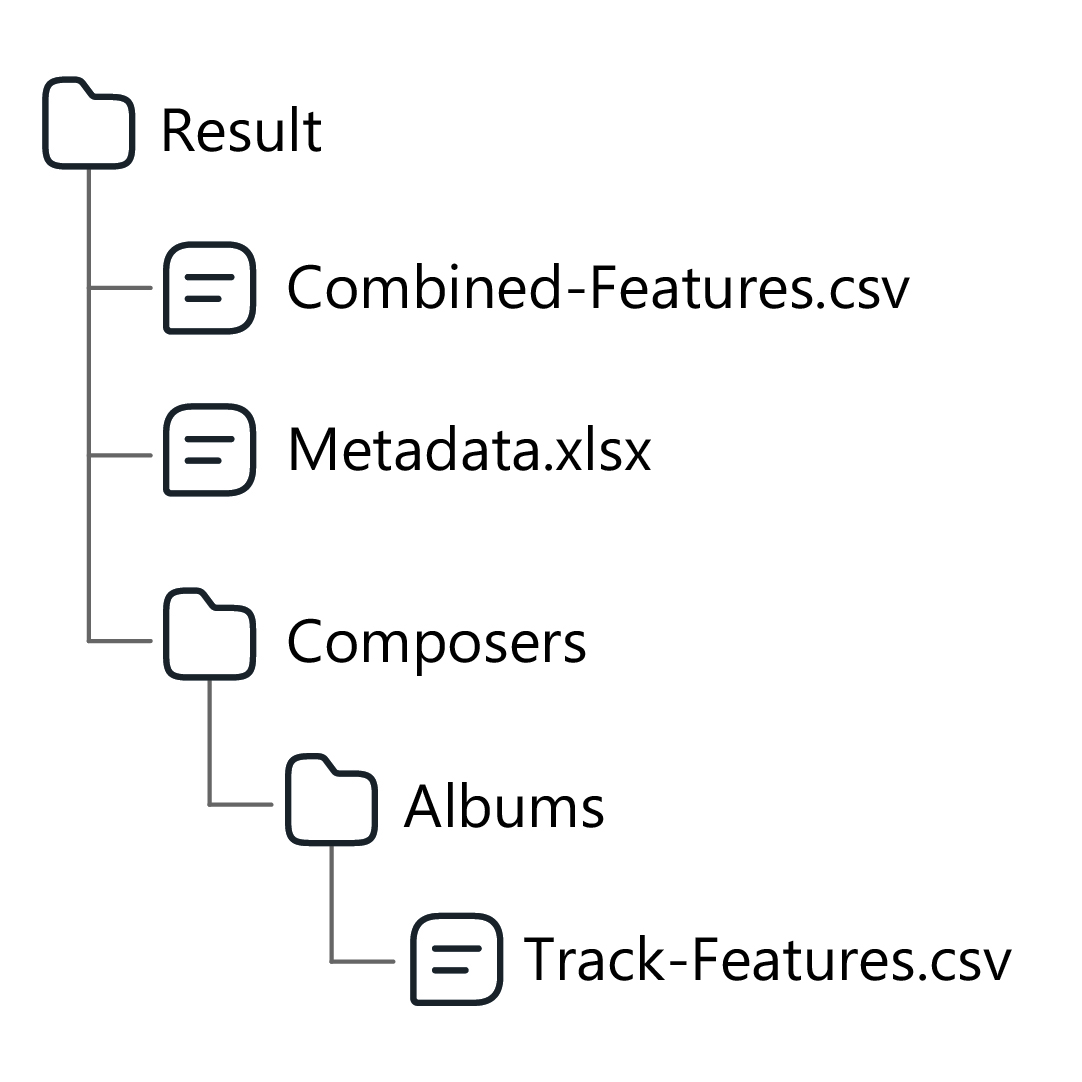}}
    \caption{The folder and file structure of the PPC corpus.}
    \label{fig: structure}
  \end{minipage}
\end{figure}

\subsection{Repository} \label{sec: Repository}
After reviewing and comparing various references, we decided to share the obtained data through Harvard Dataverse(https://www.dataverse.harvard.edu/) for public access. Easy and secure access to Persian piano data was one of our main considerations. With advanced backup systems, multiple servers, and a user-friendly interface, we chose this repository.
The data obtained from the feature extraction of pieces is stored in CSV files and is accessible in 2 separate parts :

In the basic state with a folder structure, the main "Result" folder contains subfolders with the names of composers. Each composer's folder includes further subfolders with the names of their albums, and finally, inside the album folders, CSV files with the titles of the pieces are stored.
In the second format, for easy access, all of the files are gathered in a combined CSV file. These files provide a structured format for researchers and analysts to access and utilize the feature data. 
Finally, alongside these feature files, an Excel file containing a list of all pieces with complete metadata has been provided. This file includes summarized tables next to the main table. The folder and file structure of the proposed corpus is illustrated in figure \ref{fig: structure}.

\section{Evaluation} \label{sec: Evaluation}
The proposed corpus consists of a substantial number of pieces, which, in comparison to other examples mentioned in section \ref{sec: Related works}, is the largest. Furthermore, this volume of information is solely based on one instrument, while in all other cases, a minimum of two instruments have been examined. The presence of complete metadata alongside the data further enhances the completeness of the PPC corpus, making it more comprehensive than the mentioned cases in section \ref{sec: Related works}.
A brief comparison of the cases mentioned is visible in Table \ref{tab:dataset-info}.
\begin{table}
  \centering
  \begin{tabular}{|l|r|r|l|l|}
    \hline
    Corpus & Number of Tracks & Total Length(Hours) & Instruments & Metadata \\
    \hline
    \textbf{PPC} & \textbf{2022} & \textbf{140} & \textbf{Piano} & \textbf{Included} \\
    KDC & 213 & 2 & Ney, Setar & Included \\
    \cite{gujsa98108} & 1250 & - & Zehi, Zakhme, Tar, Setar & Not included \\
    Nava & 1786 & 55 & Tar, Santoor, Setar, Kamancheh, Ney & Partial included \\
    Meter2800 & 2800 & 23 & Multi-instrument & Included \\
    MICM & 3311 & - & Straw Instrument, Violin & Not included\\
    USPOP2002 & 8764 & 597 & Multi-instrument & Partial included\\
    LCIM & 2185 & 2185 & Multi-instrument & Partial included\\
    \hline
  \end{tabular}
  \caption{corpus comparison}
  \label{tab:dataset-info}
\end{table}

\section{Conclusion} \label{sec: Conclusion}
In conclusion, the corpus represents a significant step towards addressing the dearth of comprehensive data in the realm of Persian music, particularly concerning the Persian piano. By creating a substantial music corpus and providing rich metadata, the PPC corpus serves as a valuable resource for researchers and analysts in the field of music information retrieval, acoustic corpusis, and genre classification. Furthermore, the project's contributions go beyond data collection, as it offers opportunities to delve deeper into the study of Persian music and its cultural significance. The completion of this project holds the potential to enrich the understanding of Iranian music and open new horizons for artists and researchers alike. In the next steps, we will aim to process this data extensively to provide systems for instrument recognition on the piano, utilizing the full potential of this corpus for advancing research. Additionally, alongside research, we will strive to complete and present subsequent versions of this corpus.

\bibliographystyle{plain}  
\bibliography{references}

\begin{thebibliography}{10}

\bibitem{gujsa98108}
Mahmood Abbası~Layegh, Siamak Haghıpour, and Yazdan Najafı~Sarem.
\newblock Classification of the radif of mirza abdollah a canonic repertoire of persian music using svm method.
\newblock {\em Gazi University Journal of Science Part A: Engineering and Innovation}, 1(4):57 -- 66, 2013.

\bibitem{ABIMBOLA2023109736}
Jeremiah Abimbola, Daniel Kostrzewa, and Pawel Kasprowski.
\newblock Meter2800: A novel dataset for music time signature detection.
\newblock {\em Data in Brief}, page 109736, 2023.

\bibitem{https://doi.org/10.13140/rg.2.2.18688.89602}
Shahla~Rezezadeh Azar, Ali Ahmadi, Saber MalekzadeH, and Maryam Samami.
\newblock Instrument-independent dastgah recognition of iranian classical music using azarnet.
\newblock {\em Unpublished}, 2018.

\bibitem{Nava}
B.~Baba~Ali, A.~Gorgan~Mohammadi, and A.~Faraji~Dizaji.
\newblock Nava: A persian traditional music database for the dastgah and instrument recognition tasks.
\newblock {\em Advanced Signal Processing}, 3(2):125--134, 2019.

\bibitem{9050082}
Azam Bastanfard and Dariush Amirkhani.
\newblock Detect hidden message in reverse timestamp in farsi.
\newblock In {\em 2020 25th International Computer Conference, Computer Society of Iran (CSICC)}, pages 1--8, 2020.

\bibitem{9349583}
Azam Bastanfard, Dariush Amirkhani, and Sadegh Naderi.
\newblock A singing voice separation method from persian music based on pitch detection methods.
\newblock In {\em 2020 6th Iranian Conference on Signal Processing and Intelligent Systems (ICSPIS)}, pages 1--7, 2020.

\bibitem{10.1007/978-3-642-11301-7_30}
Azam Bastanfard, Maryam Fazel, Alireza~Abdi Kelishami, and Mohammad Aghaahmadi.
\newblock The persian linguistic based audio-visual data corpus, ava ii, considering coarticulation.
\newblock In Susanne Boll, Qi~Tian, Lei Zhang, Zili Zhang, and Yi-Ping~Phoebe Chen, editors, {\em Advances in Multimedia Modeling}, pages 284--294, Berlin, Heidelberg, 2010. Springer Berlin Heidelberg.

\bibitem{10.1162/014892604323112257}
Adam Berenzweig, Beth Logan, Daniel P.~W. Ellis, and Brian P.~W. Whitman.
\newblock A large-scale evaluation of acoustic and subjective music-similarity measures.
\newblock {\em Comput. Music J.}, 28(2):63–76, jun 2004.

\bibitem{caton_1975}
Margaret~L. Caton.
\newblock Ella zonis, classical persian music: An introduction (cambridge, mass.: Harvard university press, 1973). pp. 233.
\newblock {\em International Journal of Middle East Studies}, 6(1):122–122, 1975.

\bibitem{chen-2023}
L.~Chen.
\newblock {Research on the value and strategy of integrating music education and national music Culture}.
\newblock {\em Journal of contemporary educational research}, 7(9):14--19, 9 2023.

\bibitem{defferrard-2017}
Michaël Defferrard, Kirell Benzi, Pierre Vandergheynst, and Xavier Bresson.
\newblock {FMA: a dataset for music Analysis.}
\newblock {\em arXiv (Cornell University)}, pages 316--323, 1 2017.

\bibitem{ebrat2022iranian}
Danial Ebrat, Farzad Didehvar, and Milad Dadgar.
\newblock Iranian modal music (dastgah) detection using deep neural networks, 2022.

\bibitem{Santoor1}
Peyman Heydarian and David Bainbridge.
\newblock Dastg\`{a}h recognition in iranian music: Different features and optimized parameters.
\newblock In {\em 6th International Conference on Digital Libraries for Musicology}, DLfM '19, page 53–57, New York, NY, USA, 2019. Association for Computing Machinery.

\bibitem{Heydarian2005ADF}
Peyman Heydarian and Joshua~D. Reiss.
\newblock A database for persian music.
\newblock In {\em A DATABASE FOR PERSIAN MUSIC}, 2005.

\bibitem{jitendra-2020}
Mukkamala S. N.~V. Jitendra.
\newblock {A Review: Music Feature Extraction from an Audio Signal}.
\newblock {\em International journal of advanced trends in computer science and engineering}, 4 2020.

\bibitem{Kassler1966TowardMI}
Michael Kassler.
\newblock Toward musical information retrieval.
\newblock {\em Perspectives of New Music}, 4:59, 1966.

\bibitem{7288794}
Mohammad Keshtkar and Azam Bastanfard.
\newblock Determining the best proportion of music genre to be played in a radio program.
\newblock In {\em 2015 7th Conference on Information and Knowledge Technology (IKT)}, pages 1--7, 2015.

\bibitem{10.1145/2542205.2542206}
Peter Knees and Markus Schedl.
\newblock A survey of music similarity and recommendation from music context data.
\newblock {\em ACM Trans. Multimedia Comput. Commun. Appl.}, 10(1), dec 2013.

\bibitem{Magna}
Edith Law, Kris West, Michael Mandel, Mert Bay, and J.~Downie.
\newblock Evaluation of algorithms using games: The case of music tagging.
\newblock In {\em Evaluation of Algorithms Using Games: The Case of Music Tagging.}, pages 387--392, 01 2009.

\bibitem{10.1145/860435.860487}
Tao Li, Mitsunori Ogihara, and Qi~Li.
\newblock A comparative study on content-based music genre classification.
\newblock In {\em Proceedings of the 26th Annual International ACM SIGIR Conference on Research and Development in Informaion Retrieval}, SIGIR '03, page 282–289, New York, NY, USA, 2003. Association for Computing Machinery.

\bibitem{malekzadeh-2019}
Saber Malekzadeh, Maryam Samami, Shahla~Rezazadeh Azar, and Maryam Rayegan.
\newblock {Classical Music Generation in Distinct Dastgahs with AlimNet ACGAN}.
\newblock {\em arXiv (Cornell University)}, 1 2019.

\bibitem{mcfee_2023_8252662}
others McFee, Brian.
\newblock librosa/librosa: 0.10.1, August 2023.
\newblock Full author list available at \url{https://doi.org/10.5281/zenodo.8252662}.

\bibitem{african}
Kyama~M. Mugambi.
\newblock Music on mission: Integrated perspectives on the shared song.
\newblock {\em Missiology}, 0(0):00918296231207786, 0.

\bibitem{babak_nikzat_2022_7316660}
Babak Nikzat and Rafael~Caro Repetto.
\newblock {KDC: an open corpus for computational research of dastgāhi music}.
\newblock In {\em {Proceedings of the 23rd International Society for Music Information Retrieval Conference}}, pages 321--328. ISMIR, nov 2022.

\bibitem{10.1162/014892604323112248}
Elias Pampalk, Simon Dixon, and Gerhard Widmer.
\newblock {Exploring Music Collections by Browsing Different Views}.
\newblock {\em Computer Music Journal}, 28(2):49--62, 06 2004.

\bibitem{pingle-2023}
Yogesh Pingle and L.~K. Ragha.
\newblock {An in-depth analysis of music structure and its effects on human body for music therapy}.
\newblock {\em Multimedia Tools and Applications}, 10 2023.

\bibitem{DVN/YY7SVD_2023}
Parsa Rasouli and Azam Bastanfard.
\newblock {The Persian Piano Corpus}, 2023.

\bibitem{rho-2011}
Seungmin Rho, Seheon Song, Yunyoung Nam, Eenjun Hwang, and Minkoo Kim.
\newblock {Implementing situation-aware and user-adaptive music recommendation service in semantic web and real-time multimedia computing environment}.
\newblock {\em Multimedia Tools and Applications}, 65(2):259--282, 5 2011.

\bibitem{INR-042}
Markus Schedl, Emilia Gómez, and Julián Urbano.
\newblock Music information retrieval: Recent developments and applications.
\newblock {\em Foundations and Trends® in Information Retrieval}, 8(2-3):127--261, 2014.

\bibitem{5407527}
M.~Hassan Shirali-Shahreza and Sajad Shirali-Shahreza.
\newblock Large corpus of iranian music.
\newblock In {\em 2009 IEEE International Symposium on Signal Processing and Information Technology (ISSPIT)}, pages 568--573, 2009.

\bibitem{simms2012mohammad}
R.~Simms and A.~Koushkani.
\newblock {\em Mohammad Reza Shajarian’s Avaz in Iran and Beyond, 1979–2010}.
\newblock Lexington Books, 2012.

\bibitem{tzanetakis-2002}
George Tzanetakis and Perry~R. Cook.
\newblock {Musical genre classification of audio signals}.
\newblock {\em IEEE Transactions on Speech and Audio Processing}, 10(5):293--302, 7 2002.

\bibitem{Vafaeian}
Amir Vafaeian, Keivan Borna, Hamed Sajedi, Dariush Alimohammadi, and Pouya~and Sarai.
\newblock Automatic identification and classification of the iranian traditional music scales (dastgāh) and melody models (gusheh): Analytical and comparative review on conducted research.
\newblock {\em Human Information Interaction}, 5(2), 2018.

\bibitem{ozer-2023}
Yigitcan Özer, Simon Schwär, Vlora Arifi-Müller, Jeremy Lawrence, Emre Sen, and Meinard Müller.
\newblock Piano concerto dataset (pcd): A multitrack dataset of piano concertos.
\newblock {\em Transactions of the International Society for Music Information Retrieval}, Sep 2023.

\end{thebibliography}

\end{document}